\documentclass[aps,pre,twocolumn,showpacs]{revtex4-1}

\usepackage{amssymb,amsfonts,amsmath}
\usepackage{bm}
\usepackage{graphicx}

\newcommand{\ee}{\ensuremath{\mathrm{e}}}
\newcommand{\ii}{\ensuremath{\mathrm{i}}}
\newcommand{\abs}[1]{\ensuremath{\left\vert #1 \right\vert}}

\newcommand{\eto}[1]{\ensuremath{\,{\ee}^{#1}}}
\renewcommand{\vec}[1]{{\ensuremath{\boldsymbol{#1}}}}
\newcommand{\dd}{\ensuremath{\mathrm{d}}}
\newcommand{\identity}{\ensuremath{\vec{1}}}
\newcommand{\firstderiv}[2]{\ensuremath{\frac{\mathrm{d}#1}{\mathrm{d}#2}}}
\newcommand{\qmnorm}[1]{\ensuremath{\left\langle #1\middle\vert #1 \right\rangle}}
\newcommand{\dH}{\ensuremath{{d_\mathrm{H}}}} 
\newcommand{\Ws}{\ensuremath{\mathcal{W_\mathrm{s}}}} 
\DeclareMathOperator{\Real}{Re}
\DeclareMathOperator{\Imag}{Im}

\begin{document}

\title{Fractal Weyl law for three-dimensional chaotic hard-sphere scattering systems}

\author{Alexander Ebersp\"acher}
\author{J\"org Main}
\author{G\"unter Wunner}

\affiliation{Institut f\"ur Theoretische Physik 1, Universit\"at Stuttgart, 70550 Stuttgart, Germany}

\date{\today}

\begin{abstract}
The fractal Weyl law connects the asymptotic level number with the 
fractal dimension of the chaotic repeller. We provide the first test 
for the fractal Weyl law for a three-dimensional open scattering 
system. For the four-sphere billiard, we investigate the chaotic 
repeller and discuss the semiclassical quantization of the system by 
the method of cycle expansion with symmetry decomposition. We test 
the fractal Weyl law for various symmetry subspaces and 
sphere-to-sphere separations.
\end{abstract}

\pacs{05.45.Mt, 03.65.Sq, 42.25.−p}

\maketitle

\section{Introduction}
The asymptotic eigenvalue distribution of partial differential 
equations such as the Schr\"odinger equation for free particles or 
the one-dimensional Helmholtz equation for sound waves has been of 
interest as early as from 1912 on when Hermann Weyl and Richard Courant 
first studied the problem \cite{Weyl1912a, Courant1920a}. They
found expressions for the asymptotic level number $N(k)$ in 
closed systems to be proportional to $k^{d}$, with $d$ the spatial 
dimension of the system. The so called ``Weyl law'', which has been 
well known from then on \cite{Baltes1976a}, states that for \emph 
{closed} quantum systems, every accessible Planck cell in phase 
space is occupied by one quantum state. A generalization to chaotic 
\emph{open} systems, where complex resonances $k_{n}=\bar{k}_{n} - 
\Gamma_{n}/2$ with mean energies $k_{n}$ and lifetimes $\Gamma_{n}$ 
replace real eigenvalues $k$, has been proposed in the 1990s \cite
{Sjoestrand1990a, Zworski1999a}. The number of resonances
\begin{align}
N(k) = \left\lbrace k_n\!: \; \Real(k_n) \leq k;\, \Imag(k_n) > -C \right\rbrace
\end{align}
inside a rectangle in the complex plane defined by the energy $k$ 
and the strip width $C$ is conjectured to be proportional to 
$k^{\alpha}$ with the exponent 
\begin{align}
\alpha = \frac{D + 1}{2}
\end{align}
being related to the non-integer fractal dimension $D$ of the \emph 
{chaotic repeller}. The number $\alpha$ takes the role of the 
effective number of degrees of freedom. The repeller is the set of 
all classical trajectories that stay trapped for $t \to +\infty$ or 
$t \to -\infty$. Considering only the stable manifold $\Ws$ of 
trapped trajectories for $t\to +\infty$ in a suitable Poincar\'e 
surface of section, the fractal Weyl law reads \cite{Lu2003a}
\begin{align}
N(k) \propto k^{\dH + 1} \, ,
\label{eq:WeylLaw}
\end{align}
with $\dH$ the Hausdorff dimension of the chaotic repeller's stable 
manifold.

The fractal Weyl law \eqref{eq:WeylLaw} has been investigated for 
various two-dimensional systems, e.g. a triple Gaussian potential 
\cite{Lin2002a}, the three-disk billiard \cite{Lu2003a}, an optical 
microstadium resonator \cite{Wiersig2008a} and a modified 
H\'enon-Heiles potential \cite{Ramilowski2009a} as well as for quantum 
maps, e.g. the kicked rotator \cite{Shepelyansky2008a, Kopp2010a}. 
The systems under consideration so far have all been at most 
two-dimensional. We provide a first investigation of a 
three-dimensional system in this paper.

\begin{figure}
\includegraphics[width=0.7\columnwidth]{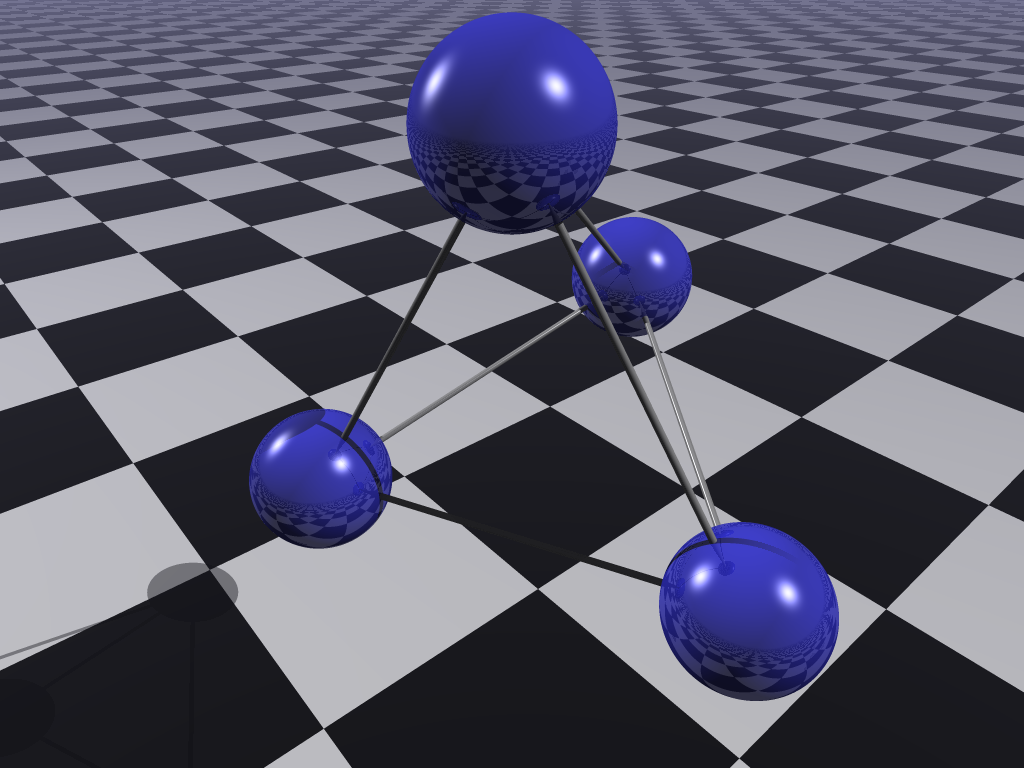}
\caption{
The four-sphere billiard. Four spheres of equal radius $R$ are 
located on the vertices of an equilateral tetrahedron (indicated by bars) 
with edge length $d$. Shown is the case of $d/R={6}$.}
\label{fig:4sphereVisu}
\end{figure}

The problem under consideration is scattering in the four-sphere billiard. 
This system is characterized by 
four spheres of the same radius $R$ located on the vertices of an 
equilateral tetrahedron with edge length $d$ as visualized in Fig.\
\ref{fig:4sphereVisu}.
The relevant configuration parameter is the ratio $d/R$. Different from 
e.g. the two-dimensional three-disk billiard, the four-sphere billiard 
is an open system even for the case of touching spheres at $d/R=2$.

The paper is organized as follows.
The fractal dimension of the repeller is computed in Sec.\ \ref{chap:Gauging}.
Periodic-orbit theory and cycle expansion methods are applied in 
Sec.\ \ref{chap:Semiclass} to obtain semiclassical resonance spectra.
The classical escape rate is determined in Sec.\ \ref{sec:escape}.
Results which allow for a test of the fractal Weyl law are presented in
Sec.\ \ref{chap:Results}.
Concluding remarks are given in Sec.\ \ref{sec:conclusion}.

\section{Gauging the repeller}
\label{chap:Gauging}
As the fractal dimension $D$ of the chaotic repeller enters into the 
fractal Weyl law \eqref{eq:WeylLaw}, it is crucial to determine $D$ 
accurately. The fractal structures in the four-sphere billiard have 
already been studied experimentally and theoretically \cite
{Sweet1999a, Sweet2001a, Chen1990a, Motter2000a}, however, all 
investigations so far have been limited to small values of the 
configuration parameter $d/R \lesssim 2.5$. In this paper the 
fractal dimension of the repeller is determined accurately for a 
wide range of the parameter $d/R$.

\subsection{Fractal repeller}
\label{subsec:FractRepller}
\begin{figure}
\includegraphics[width=0.9\columnwidth]{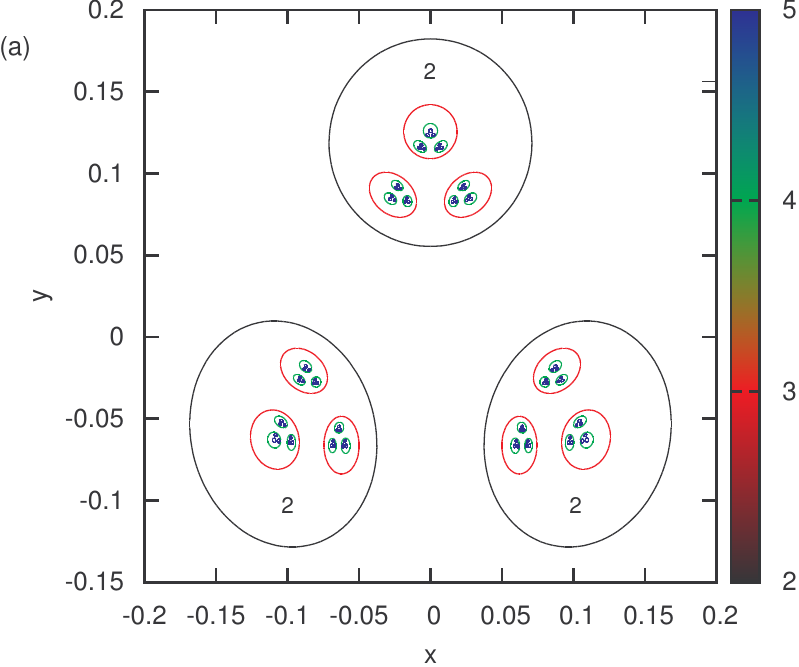}
\includegraphics[width=0.9\columnwidth]{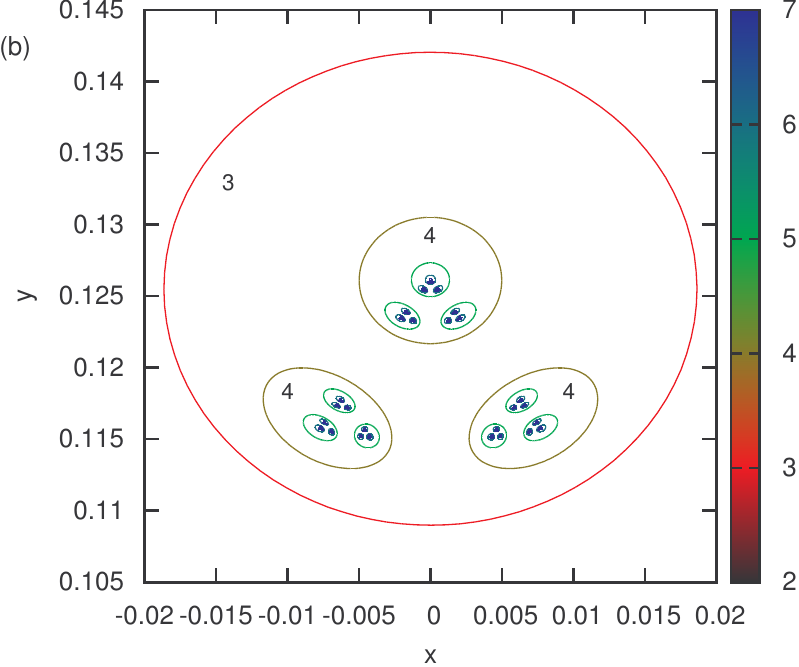}
\caption{
(a) Time-delay functions in the range $2 \le T \le 5$ for $d/R=3$  and 
(b) a magnification thereof in the range $2 \le T \le 7$.  The time-delay 
functions are drawn as functions of the coordinates $x,y$ in the surface of 
section.  The colors indicate the value of $T$ 
and for clear identification some regions are explicitly labeled with $T$.
Only the boundaries of individual regions are drawn; 
initial conditions chosen within the regions experience the same 
number of reflections. The self-similarity suggests that the stable 
manifold is a fractal set.}
\label{fig:TD4diskIllu}
\end{figure}

The stable manifold $\Ws$ of the chaotic repeller is a fractal in 
phase space. The time spent in the scattering system can be measured 
by the time-delay function $T$ counting the number of 
reflections experienced by a trajectory. Choosing initial conditions 
in a plane parallel to the plane spanned by three of the spheres, it 
is possible to iterate trajectories entering the scattering system 
such that the fourth sphere is visited first. For large ratios $d/R$, 
the boundary to the region that contains those initial conditions 
is the projection of the fourth sphere onto the plane from which the 
trajectories are iterated. As there are three distinct possibilities 
to visit the next sphere, there are three regions of higher values of 
$T$ inside this circle. Repeating this line of argument for any region 
belonging to a given visitation sequence, the structure  of the fractal 
repeller can be understood. Figure \ref{fig:TD4diskIllu} illustrates 
the fractal structure.

In principle, the Hausdorff dimension $\dH$ can be calculated from 
box-counting. This procedure, however, is not suited for the 
billiards under consideration as it requires iteration of a vast 
amount of initial conditions on a grid. The regions of high values of
$T$ that exhibit fractal properties may not be resolved with acceptable 
computational effort. A method better suited to billiards is introduced 
below.

\subsection{Estimating $\dH$ through Hausdorff sums}
Finite numerical precision and finite computing time available 
prevent the determination of initial conditions that lead to trapped 
orbits. Instead, it is possible to estimate the Hausdorff dimension 
from regions of finite $T$. We introduce the auxiliary quantity 
$A^{(i)}_n$ that denotes the area of the $n$-th region of initial 
conditions with $T \ge i$ reflections in the surface of section, and 
define the quantities
\begin{align}
K^{(i)}(s) := \sum\limits_{n} \left( A^{(i)}_n \right)^s
\label{eq:HausdorffSums}
\end{align}
which will be called Hausdorff sums below. These sums have 
the following properties \cite{Falconer1985a, Zaslavsky2005a}:
\begin{align}
\lim\limits_{i\to\infty} K^{(i)}(s) = \left\lbrace
\begin{array}{ll}
  \infty                      &   \textrm{for } 0 \leq s < \dH \\
  \textrm{const.}>0 \quad    &   \textrm{for } s = \dH\\
  0                           &   \textrm{for } \dH < s < \infty
\end{array}
 \right.
\end{align}
The property for $s=\dH$ stems from the fact that the Hausdorff sums 
by definition are smooth functions of the variable $s$. This allows one
to estimate the Hausdorff dimension $\dH$ by intersecting 
$K^{(i)}(s)$ for different $i$ \cite{Chen1990a, Motter2000a}.

Existing methods of estimating the repeller's fractal dimension have 
been confined to a narrow range of the parameter $d/R$, in particular
to the case of almost-touching spheres \cite{Chen1990a, 
Motter2000a}. The algorithm discussed in the following allows for 
calculations in a wider range of $d/R$. To estimate the stable 
manifold $\Ws$ in the surface of initial conditions, it will be 
necessary to accurately find boundaries of regions of a given 
visitation sequence of the spheres and reflection count. Once both 
a point inside the region in question and a point outside are known, 
interval bisection may be used to compute the boundary point on the 
line connecting both points. The bisection condition uses the 
visitation sequence and the reflection count $T$, i.e.\ the length 
of the symbolic code.

\subsubsection{Finding regions}
The algorithm used in this paper relies on the structure of the 
time-delay functions discussed in Sec.\ \ref{subsec:FractRepller}. 
In the calculations, the following assumptions are made:
\begin{itemize}
\item Regions of a specific order of visits with the scatterers
described by the symbolic code are non-overlapping.
\item Within a region of a given order of visits, there are exactly 
three more regions each corresponding to additional visits at one of 
the three other spheres.
\end{itemize}
Both conditions may be violated for $d/R$ close to $2$, i.e.\ the 
case of almost touching spheres. We find the assumptions to be 
fulfilled for regions with $T \geq 2$ and configurations $d/R\gtrsim 2.5$.

All steps of the procedure are based on the Poincar\'e surface of 
section chosen such that iteration starts from a plane parallel to 
the plane spanned by the three closest spheres. The velocities are 
chosen parallel to the $z$-axis such that the uppermost sphere is 
visited first. Let us assume that regions with $T_\mathrm{min} \le T \le 
T_\mathrm{max}$ are sufficient for an estimation of $\dH$. Under 
this assumption, it is possible to find regions approximating the 
repeller with the following procedure.

In the a first step, the projection of the sphere visited first onto 
the surface of section is determined. For small $d/R$, this region 
is a circle, for larger $d/R$, projections of the other spheres may 
be cut out of the circle. This is done by randomly choosing a point 
in the surface of section and an interval bisection between this 
point and points equally distributed on a large circle fully 
containing the projection of sphere $4$. By assumption, inside this 
region there are three other regions with $T=2$. The corresponding 
visitation sequences differ in the second character. Once a point 
inside each of the regions with $T=2$ is found, in a second step, 
polygonal chains forming boundaries to each of the new distinct 
regions are calculated. This procedure is iterated until all desired 
regions corresponding to $T_\mathrm{min} \le T \le T_\mathrm{max}$ 
have been found.

\subsubsection{Areas from polygonal chains}
One possibility to store the boundaries is by keeping a polygonal 
chain. As in this procedure the number $N_\mathrm{regions}$ of 
regions grows exponentially with $N_\mathrm{regions} = 3^{T-1}$, this 
way of data storage is memory-expensive. However, the areas enclosed 
by the polygons are easily calculated using numerical quadrature of 
the area given by
\begin{align}
A = \frac{1}{2} \int\limits_{0}^{2\pi} r^{2}(\varphi) \,\dd \varphi\, ,
\end{align}
with $r(\varphi)$ the distance of the boundary point from the 
``midpoint'' of the region. A fairly low number of supporting points 
has proved to be sufficient for very high precision. All 
calculations have been performed with $101$ supporting points.

\subsubsection{Areas from ellipses}
An alternative to the memory-expensive storage of polygonal chains 
is to approximate the boundary by an ellipse described by the 
polynomial
\begin{align}
 a_1 x^2 + a_2 y^2 + a_3 xy + a_4 x + a_5 y = 1 \, .
\label{eq:EllipPoly}
\end{align}
For five or more known points $(x_i, y_i)$ of the boundary the 
coefficients $a_1,\dots,a_5$ can be determined from a linear 
least-squares fit. The semi-major axes $a,b$, the center shift $(x_0, 
y_0)$ as well as the rotation angle $\varphi$ of the ellipses can be 
easily extracted from the coefficients of the polynomial in Eq.\ 
\eqref{eq:EllipPoly}.

Fitted ellipses allow to improve accuracy as it is now possible to 
shift the regions' ``midpoints'' used in the construction of the 
polygonal chains to the midpoint of the ellipses. All bisections for 
the polygonal chain boundaries are repeated in such a way that all 
lines connecting the ellipse's midpoints and the boundary points 
intersect at identical angles. This will be beneficial for the 
quadrature of the areas entering into the Hausdorff sums.

Once all desired boundaries have been calculated, the fractal 
dimension $\dH$ can be estimated. To build the Hausdorff sums \eqref
{eq:HausdorffSums}, the areas enclosed in the individual regions 
have to be known. From the ellipses fitted to the boundaries, the 
area $A$ is trivially given by
\begin{align}
A=\pi a b\, ,
\end{align}
where $a$ and $b$ are the semi-major axes.

\begin{figure}
\includegraphics[width=0.9\columnwidth]{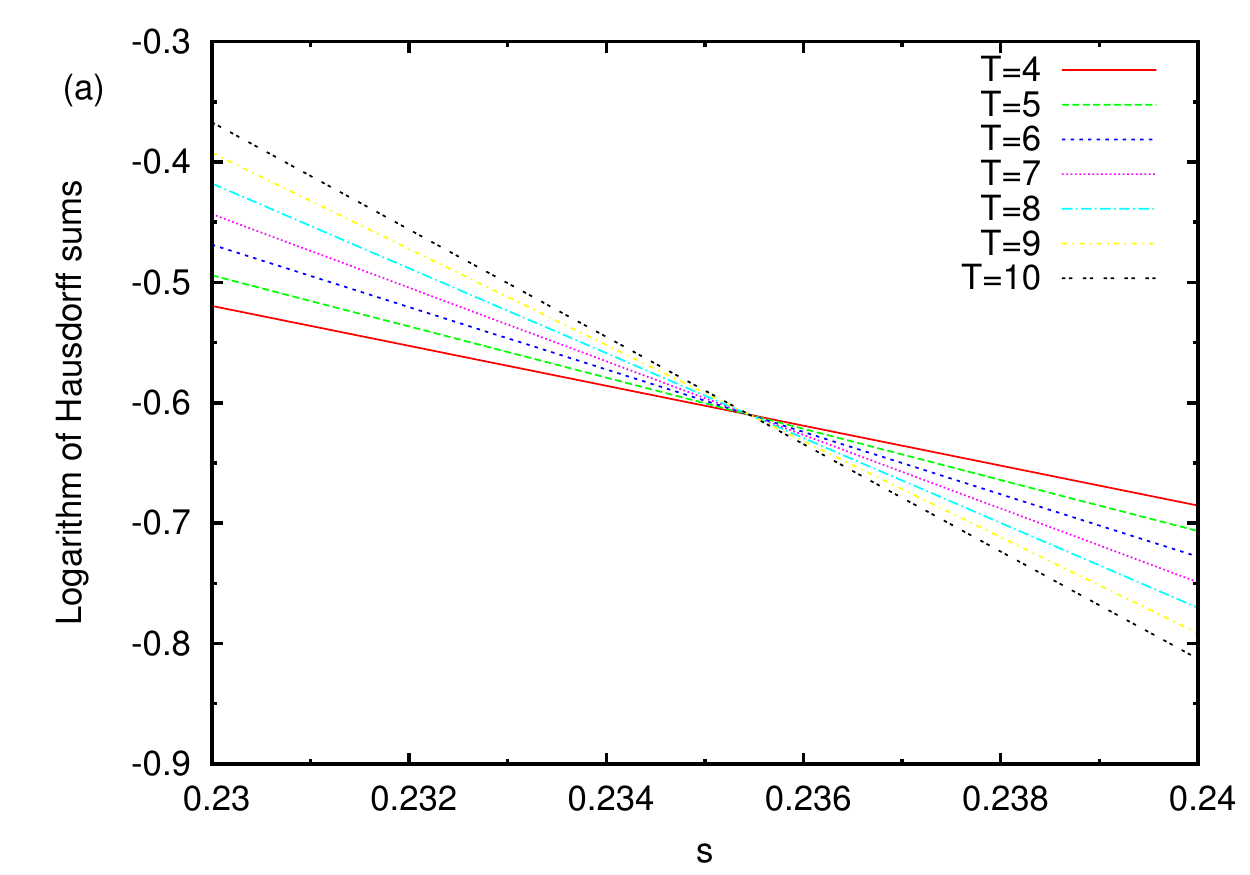}
\includegraphics[width=0.9\columnwidth]{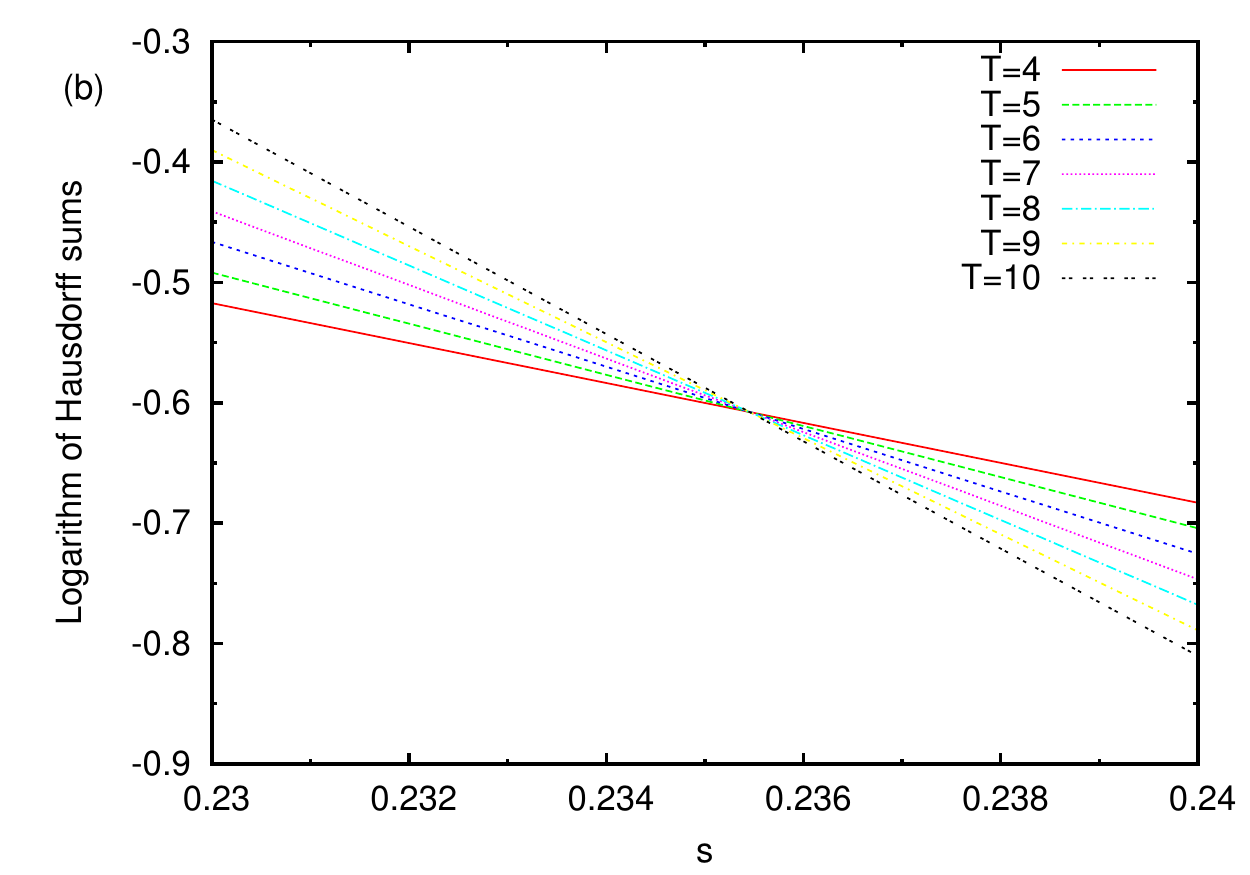}
\caption{
Intersected Hausdorff sums $K^{(i)}(s)$ for various 
reflection numbers $i=T$ calculated from (a) a polygonal chain with $101$ 
supporting points and (b) ellipses fitted to polygonal chains for $d/R=6$. 
As can clearly be seen, the intersection points for all shown curves 
agree perfectly. For this reason, the Hausdorff dimensions can be 
determined to a precision of at least $4$ significant digits.}
\label{fig:SampleHDSum}
\end{figure}
\begin{figure}[t]
\includegraphics[width=0.9\columnwidth]{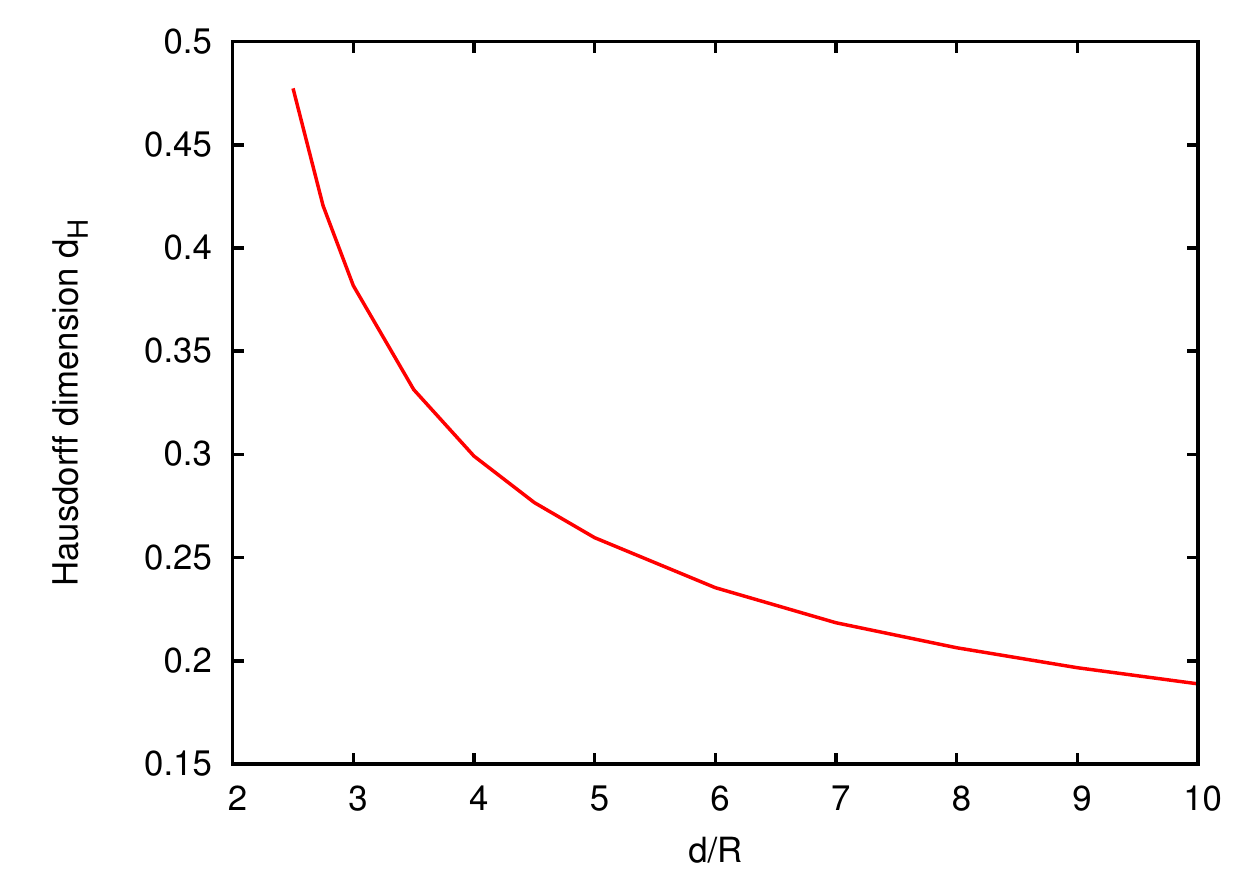}
\caption{
Hausdorff dimension $\dH$ of the stable manifold $\Ws$ as function
of the ratio $d/R$. All data points have been obtained by intersecting 
Hausdorff sums as demonstrated in Fig.~\ref{fig:SampleHDSum}.}
\label{fig:dHResults}
\end{figure}

Calculations have been performed for $d/R = 2.5$ to $d/R = 10$.
Sample plots for intersected Hausdorff sums $K^{(i)}(s)$ are shown
in Fig.~\ref{fig:SampleHDSum}. Results for the Hausdorff
dimension $\dH$ are compiled in Fig.~\ref{fig:dHResults} and in
Table~\ref{tab:dH}. The calculations using polygonal chains
agree up to four decimal digits with the calculations using fitted
ellipses.
\begin{table}
\caption{Numerical values of the Hausdorff dimensions $\dH$ of the
stable manifold $\Ws$ for various configuration parameters $d/R$.
All decimal digits are significant.}
\begin{tabular}{l| c c c c c c c}
 $d/R$  & $2.5$ & $3$ & $4$ & $5$ & $6$ & $8$ & $10$\\
 $\dH$ & $0.4774$ & $0.3818$ & $0.2992$ & $0.2596$ & $0.2354$ & $0.2063$ &
 $0.1888$
\end{tabular}
\label{tab:dH}
\end{table}

Figure~\ref{fig:dHResults} clearly shows that with decreasing $d/R$ 
the intersection of the stable manifold $\Ws$ with the Poincar\'e 
surface of section fills the plane denser. The repeller's dimension 
$\dH$ thus increases as the tetrahedron gets packed more densely.

In summary, the method presented here establishes a fast and very 
precise method of gauging the repeller. Though the assumptions are 
quite strong, they hold over a wide range of the ratio $d/R$.

\section{Semiclassical resonances}
\label{chap:Semiclass}
Studying a billiard system in a purely quantum mechanical fashion
turns out to be intricate since -- although particles move freely in
between the scatterers -- it is a demanding task to find
wavefunctions that vanish on all scatterer's boundaries
simultaneously. For attempts on $N$-sphere scattering systems in
three dimensions, see \cite{Henseler1997a}; for the two-dimensional
three-disk scattering system see \cite{Gaspard1989c}. Results and
comparisons of methods for the four-sphere scattering system are
presented in \cite{Main2004a}. The techniques of \emph{semiclassical
quantization} presented below are better suited for billiard
systems.

As one of the great achievements of semiclassical physics, 
Gutzwiller's trace formula provides a mean of 
quantizing a system via periodic orbits \cite{Gutzwiller90a}. Unfortunately, 
the trace formula is plagued by serious convergence problems. In chaotic 
systems, the number of periodic orbits typically grows exponentially 
with length $l$, and this growth usually cannot be compensated by the 
decrease of the amplitude factors.
A method of improving the convergence ideally suited for billiard 
systems is based on the Gutzwiller-Voros zeta function.

The logarithmic derivative of the function
\begin{align}
Z(k) = \prod\limits_{n}(k-k_n)
\end{align}
with the quantized wavenumbers $k_n$ of a billiard system yields the 
density of states
\begin{align}
g(k) = -\frac{1}{\pi} \Imag \firstderiv{}{k} \ln Z(k)
 =-\frac{1}{\pi} \Imag \sum\limits_{n} \frac{1}{k-k_n + \ii\epsilon}\, .
\label{eq:gk}
\end{align}
Voros \cite{Voros1988a} proposed a semiclassical formulation of $Z(k)$ 
which reads for billiard systems
\begin{align}
Z_\mathrm{GV}(k) = \exp\left( - \sum\limits_p\sum\limits_{r=1}^{\infty} 
\frac{1}{r}\frac{(-1)^{r n_p}\eto{\ii rl_pk}} {\sqrt{\abs{\det( {{\vec{M}}_p}^{r} - \identity 
)}}}\right)\, ,
\label{eq:ZGVBilliard}
\end{align}
where $l_p$ is the length of a primitive periodic orbit and $n_p$ the number
of reflections on hard wall boundaries.
The index $r$ counts the number of repetitions of a primitive periodic orbit.
The reduced monodromy matrix ${\vec{M}}_{p}$ provides information on the 
linear evolution of a small deviation from an initial condition 
belonging to a periodic orbit over one period \cite{Brack96a}. 
The eigenvalues of ${\vec{M}}_{p}$ quantify the stability of the periodic 
orbit. Due to the symplectic structure of Hamiltonian mechanics, 
the eigenvalues come in tuples 
$\lambda$, $1/\lambda$, $\lambda^*$, and $1/\lambda^*$.

For systems with symbolic dynamics such as billiards, the method of 
\emph{cycle expansion} 
\cite{Cvitanovic1988a,Cvitanovic1989a,Artuso1989a,Wirzba1999a} has 
proved to be especially successful. A cycle expansion of the Gutzwiller-Voros 
zeta function $Z_\mathrm{GV}(k)$ in Eq.\ \eqref{eq:ZGVBilliard} is 
achieved by replacing $(-1)^{r n_{p}}$ in Eq.\ \eqref
{eq:ZGVBilliard} by the term $(-z)^{r n_{p}}$ depending on the 
book-keeping variable $z$, expanding $Z_\mathrm{GV}$ as a power 
series in $z$ and then truncating the series. The highest power of 
$z$ equals the maximum cycle length $n_\mathrm{max}$ contributing to 
the cycle expansion. After truncation, $z$ has to be set to $z=1$. 
The cycle-expanded zeta function has better convergence behavior 
over the trace formulas as individual terms tend to cancel.

\subsection{The symmetry group $T_d$}
\label{subsec:SymmGroup}
The four-sphere billiard has discrete tetragonal symmetry \cite{Main2004a}. 
The associated symmetry group $T_d$ contains all symmetry operations 
that leave a regular tetrahedron invariant. In particular, there are 
the identity operation $E$, $4$ rotations $C_3$ by $2\pi/3$ around 
the axes defined by a vertex of the tetrahedron and the center of 
the facing triangular boundary surface, $4$ more rotations ${C_3}^2$ 
by $4\pi/3$ around the same axes,  $3$ rotations $C_2$ by $\pi$ 
around the axes intersecting the middle points of opposing edges, $6$
reflections $\sigma_d$ at planes perpendicular to the tetrahedron's 
edges and also containing another vertex; and, furthermore, $3$ 
permutations of the vertices $S_4$ which can be written as a 
combination of a rotation $C_4$ by $\pi/2$ and a reflection 
$\sigma_h$ at the plane perpendicular to the main rotation axis, i.e.\ 
the axes of $C_3$. Finally, the symmetry group $T_d$ also contains 
$3$ distinct three-times repeated rotary reflections ${S_4}^3$.

The character table of the symmetry group is given in
Table~\ref{tab:charTableTd}.
\begin{table}
\caption{Character table for the group $T_{d}$ \cite{Main2004a}.
The group has five irreducible representations: the one-dimensional 
representations $A_1$ and $A_2$, the two-dimensional representation 
$E$ and the two three-dimensional representations $T_1$ and $T_2$.}

\begin{tabular}{l| r r r r r}
${T_d}$ & $E$ & $8C_3$ & $3C_2$ & $6S_4$ & $6\sigma_d$ \\
\hline
$A_1$ & $1$ & $1$ & $1$ & $1$ & $1$ \\
$A_2$ & $1$ & $1$ & $1$ & $-1$ & $-1$ \\
$E$ & $2$ & $-1$ & $2$ & $0$ & $0$ \\
$T_1$ & $3$ & $0$ & $-1$ & $1$ & $-1$ \\
$T_2$ & $3$ & $0$ & $-1$ & $-1$ & $1$
\end{tabular}
\label{tab:charTableTd}
\end{table}
The symmetry group can be decomposed into $5$ invariant subspaces, 
i.e.\ the representation matrices $\vec{D}$ of $T_d$ can be 
decomposed into block-diagonal form where the diagonal elements 
contain matrices representing the group's elements. The 
representation is called ``irreducible'' if no further decomposition 
is possible.

From the character table, statements on the wavefunctions $\psi$ can 
be made. For the one-dimensional representations $A_1$ and $A_2$, 
the effect of the symmetry transformations is described by a 
multiplication with the character $\chi$. For the totally symmetric 
$A_1$ subspace, all characters are equal to $1$, therefore the
wavefunctions have the full symmetry of $T_d$, i.e.\ $\psi$ is not 
altered by any symmetry transformation. In the $A_2$ subspace, the 
wavefunction changes sign under the reflection $\sigma_d$ and 
permutation $S_4$. For representations of higher dimension, the 
effect of the symmetry transformations cannot be described in such
a simple way.

We note that repeated application of symmetry transforms may be
identical with other elements of the symmetry group, e.g.,
${\sigma_d}^2={C_2}^2={C_3}^3={S_4}^4=E$ and ${S_4}^2=C_2$.
These identities will be useful for the symmetry decomposition of
zeta functions discussed below.

\subsection{Symbolic dynamics and periodic orbits in the four-sphere scattering system}
\label{sec:symblDyn}
The method of cycle expansion requires all periodic orbits 
up to a given maximum cycle length $n_r$. In special billiards it is 
convenient to assign a \emph{symbolic code} to each periodic orbit.
For the four-sphere scattering system the symbolic dynamics and periodic 
orbits have already been introduced in Ref.\ \cite{Main2004a}.
For the convenience of the reader we briefly recapitulate the central
ideas and properties of the orbits.

In the four-sphere scattering system, periodic orbits are determined
by the periodic sequence in which the four scatterers are visited.
For most cases, any sequence corresponds to one periodic orbit. If this 
one-to-one correspondence is not given, i.e.\ if some orbits become 
unphysical because they penetrate the scatterers, one speaks of \emph 
{pruning}. This is the case for small center-to-center 
separations. In the four-sphere scattering system, the symbolic 
dynamics has been shown to be pruned for configurations with 
$d/R < 2.0482$ \cite{Main2004a}.

\subsubsection{Periodic orbits in the fundamental domain}
In systems with discrete symmetries such as the four-sphere 
billiard, which is invariant under all symmetry operations of the 
tetrahedron group $T_d$, whole classes of orbits are 
equivalent to each other, e.g., the $6$ orbits which are scattered
back and forth between two spheres can be mapped onto each other using 
symmetry operations of $T_d$.  Furthermore, cyclic permutation of the 
sequence of spheres leaves the orbits invariant. 
For these reasons, it is appropriate to use the symmetry properties to 
introduce the following short notation \cite{Main2004a}. First, define 
the plane of reflection as the plane that contains the centers of 
the last three distinct spheres visited. Then, instead of labeling 
all spheres individually, the label $0$ will be used if the orbit 
visits the last sphere once more, $1$ will indicate a visit at the 
third other sphere in the same plane of reflection, whereas the 
label $2$ will be used for a visit at the fourth sphere outside the 
plane of reflection. With this nomenclature, all symbolic codes 
containing the character $2$ are three-dimensional, whereas orbits 
corresponding to sequences of $0$ and $1$ are two-dimensional. The 
new labeling reduces the number of characters in the alphabet to 
three, i.e.\ the code is ternary. This reduction corresponds to a 
reduction of the full physical phase space $M$ to the fundamental 
domain $\tilde{M}$ from which the whole phase space can be 
reconstructed by applying the symmetry group's elements. 
Note that the symmetry reduced orbits are in general shorter than
the corresponding physical ones. Only the symmetry reduced orbits
that have the identity operation $E$ as maximum symmetry have the
same length as the corresponding physical orbits. The reduced orbits
of symmetry classes $\sigma_d$ and $C_2$ yield twice as long
physical orbits, $C_3$ orbits are three times longer than in the
physical space, and, finally, $S_4$ orbits have quadruple length.

\subsubsection{Finding periodic orbits}
Periodic orbits of the four-sphere scattering system are calculated by
varying, for a given symbolic code, the reflection points on the spheres 
until the length of the orbit takes its minimum value.
Details of the numerical periodic orbit search are described 
in \cite{Main2004a} and the computation of the monodromy matrix
is explained in Refs.\ \cite{Primack2000a,Sieber1998a}. 
Table~\ref{tab:POdata} lists the first few periodic orbits in the 
fundamental domain as well as their properties for the ratio $d/R=4$. 
\begin{table}
\caption{Primitive periodic orbits up to cycle length $n_{\tilde p} 
= 2$ for $d/R=4$. The reduced symbolic code $\tilde p$ as well as 
the symmetry $h_{\tilde p}$ of each cycle is given. Furthermore, the 
real and imaginary parts of the stability eigenvalues 
$\lambda_{\tilde p}^{(i)}$ are tabulated. All numbers have been 
rounded to five decimal digits. The shortest cycle, labeled by $0$, 
which visits two spheres in turns has ambiguous symmetry. Both the 
rotation about $\pi$, $C_2$, as well as the reflection about the 
plane perpendicular to the line connecting the sphere's center, 
$\sigma_d$, map this particular orbit onto itself.}
\begin{tabular}{ccrrrrr}
$\tilde p$ & $h_{\tilde p}$ & 
\multicolumn{1}{c} {$l_{\tilde p}$} & 
\multicolumn{1}{c} {$\Real\lambda_{\tilde p}^{(1)}$} & 
\multicolumn{1}{c} {$\Imag\lambda_{\tilde p}^{(1)}$} & 
\multicolumn{1}{c} {$\Real\lambda_{\tilde p}^{(2)}$} & 
\multicolumn{1}{c} {$\Imag\lambda_{\tilde p}^{(2)}$} \\
\hline
0 & $\sigma_d, C_2 $ & 2.00000 &  5.82843 & 0.00000 & 5.82843 &  0.00000\\
1 & $C_3$ & 2.26795 & -7.09669 &  0.00000 &  5.75443 &  0.00000 \\
2 & $S_4$&  2.31059 & -3.11111 &  5.69825 & -3.11111 & -5.69825 \\
01 & $\sigma_d$  &  4.34722& -46.21054 & 0.00000 & 32.08725 &  0.00000 \\
02 & $C_3$ & 4.35831 & -14.95013 &  35.68205 & -14.95013 &  -35.68205\\
12 & $S_4$ & 4.58593 &  43.79192 & 0.00000 & -39.51750 &  0.00000
\end{tabular}
\label{tab:POdata}
\end{table}
The table also gives the maximum symmetry operation that leaves the 
orbit invariant. In the fundamental domain, this operation 
corresponds to the operation that maps the endpoint of the orbit in 
the fundamental domain onto the starting point. Note that, for 
example, the cycle $t_0$ which visits two spheres in turns is 
periodic in the fundamental domain, but not in full physical space. 
In the full domain, the start and endpoint of the $0$-cycle are not 
identical; application of the rotation $C_2$ respectively the 
reflection $\sigma_d$ yields back the full periodic orbit.

\subsection{Discrete symmetries and cycle expansion}
In systems with discrete symmetries the full physical spectrum can 
be decomposed into spectra belonging to different representations of 
the symmetry group. The discrete symmetries lead to symmetry 
factorized zeta functions, which allow for the computation of 
quantum spectra belonging to a specific symmetry subspace. The 
symmetry decomposition of zeta functions has been elaborated by 
Cvitanovi\'c and Eckhardt \cite{Cvitanovic1993a} and examples have 
been given for the symmetry groups of various two-dimensional $N$-disk 
pinball models. Here, we present explicit results for the 
tetrahedron group $T_d$ of the three-dimensional four-sphere 
scattering system. As in \cite{Cvitanovic1993a} we first discuss the 
symmetry decomposition of the dynamical zeta function
\begin{align}
 Z(k) = \prod_p\left(1-t_p(k)\right)
\label{eq:Zdyn}
\end{align}
which is obtained from the Gutzwiller-Voros zeta function 
\eqref{eq:ZGVBilliard} with the approximation
\begin{align}
\abs{\det({{\vec{M}}_p}^r-\identity)}^{-1/2}\approx
 |\lambda_{p}^{(1)}\lambda_p^{(2)}|^{-r/2} = \ee^{-(u_p^{(1)}+u_p^{(2)})r/2}
\end{align}
and the definition
\begin{align}
t_p(k)=\ee^{\ii l_p k-(u_p^{(1)}+u_p^{(2)})/2} \; ,
\end{align}
and then generalize the results for the symmetry decomposition of the 
Gutzwiller-Voros zeta function.

In quantum mechanics, the full Hilbert space $\mathcal{H}$ of the
problem factorizes into subspaces belonging to certain irreducible
representations of the symmetry group, i.e.\ 
\begin{align}
\mathcal{H} = \mathcal{H}_{A_1} \otimes \mathcal{H}_{A_2} \otimes
\mathcal{H}_{E} \otimes \mathcal{H}_{T_1} \otimes \mathcal{H}_{T_2}\, ,
\end{align}
for the four-sphere scattering system. In \cite{Cvitanovic1993a} it is
pointed out that zeta functions can be factorized in a similar way.
The fundamental domain of phase space is sufficient for all computations, 
as the whole phase space $M$ can be obtained from the fundamental 
domain $\tilde{M}$ by
\begin{align}
M=\sum\limits_{h\in G} h \tilde{M}\, ,
\end{align}
where $G$ is the symmetry group. Evaluating traces of transfer 
operators in the fundamental domain $\tilde{M}$, this symmetry 
reduction results in \cite{Cvitanovic1993a, Artuso1989a}
\begin{align}
\left( 1 - t_p \right)^{m_p} = \det\left(\vec{1} -
\vec{D}\left({h}_{\tilde{p}}\right)t_{\tilde{p}}
\right) \, ,
\end{align}
with $m_{p}$ the multiplicity of a primitive cycle $p$.

These expressions could be evaluated using a certain explicit 
representation $\vec D_\alpha({h})$ of the group's symmetry 
operations $h$. However, this is a computationally rather demanding 
endeavor. Instead, the determinants can be expressed in terms of 
traces $\chi$ which can be read off from the symmetry 
group's character table (see Table~ \ref{tab:charTableTd}). For 
example, the expansion of $\det(\identity-\vec{D}(h)t)$ for 
dimension $d=3$ reads
\begin{align}
 \det(\identity-\vec{D}(h)t) = 1 &- \chi(h)t +
 \frac{1}{2}\left(\chi(h)^2-\chi\left(h^2\right)\right)t^2 \nonumber \\
  &+\frac{1}{6}\left(\chi(h)^3-3\chi(h)\chi\left(h^2\right)
  +2\chi\left(h^3\right)\right)t^3 \; ,
\end{align}
where the trace of $\vec{D}(h)$ is as usual denoted by $\chi(h)$.
Carrying out this procedure explicitly, one obtains the factorizations 
given in Table~\ref{tab:Factorization}.
\begin{table*}
\caption{Symmetry factorization of the zeta function $Z$ for all 
five irreducible representations of the group $T_d$. The table 
entries give the contribution of each fundamental cycle $\tilde{p}$ 
to the Euler product $Z = \prod_{\tilde{p}} (1 - t_{\tilde{p}})$. 
This factorization allows the computation of quantum spectra for 
each symmetry subspace.}
\begin{tabular}{l| c c c c c}
   & $E$ & $C_3$ & $C_2$ & $S_4$ & $\sigma_d$\\
\hline
  $A_1$ & $(1-t_{\tilde{p}})$ & $(1-t_{\tilde{p}})$ & $(1-t_{\tilde{p}})$ & $(1-t_{\tilde{p}})$ & $(1-t_{\tilde{p}})$\\
  $A_2$ & $(1-t_{\tilde{p}})$ & $(1-t_{\tilde{p}})$ & $(1-t_{\tilde{p}})$ & $(1+t_{\tilde{p}})$ & $(1+t_{\tilde{p}})$\\
  $E$ & $(1-t_{\tilde{p}})^2$ & $(1+t_{\tilde{p}}+t_{\tilde{p}}^2)$ & $(1-t_{\tilde{p}})^2$ & $(1-t_{\tilde{p}})(1+t_{\tilde{p}})$ &
  $(1-t_{\tilde{p}})(1+t_{\tilde{p}})$\\
 $T_1$ & $(1-t_{\tilde{p}})^3$ & $(1-t_{\tilde{p}})(1+t_{\tilde{p}}+t_{\tilde{p}}^2)$ & $(1-t_{\tilde{p}})(1+t_{\tilde{p}})^2$ & $(1-t_{\tilde{p}})(1+t_{\tilde{p}}^2)$ &
  $(1-t_{\tilde{p}})(1+t_{\tilde{p}})^2$\\
 $T_2$ & $(1-t_{\tilde{p}})^3$ & $(1-t_{\tilde{p}})(1+t_{\tilde{p}}+t_{\tilde{p}}^2)$ & $(1-t_{\tilde{p}})(1+t_{\tilde{p}})^2$ & $(1+t_{\tilde{p}})(1+t_{\tilde{p}}^2)$ &
  $(1+t_{\tilde{p}})(1-t_{\tilde{p}})^2$
\end{tabular}
\label{tab:Factorization}
\end{table*}
Thus, the zeta function in Eq.\ \eqref{eq:Zdyn} can be rewritten in a 
symmetry reduced version
\begin{align}
Z_\alpha =\prod\limits_{\tilde{p}} (1-\vec{D}_\alpha({h}_{\tilde{p}}) t_{\tilde{p}})
\label{eq:ZetaSymm}
\end{align}
for the subspace $\alpha$. The zeta function now depends only on the 
fundamental cycles $\tilde{p}$. By this procedure, a factorization
\begin{align}
Z(k) = \prod\limits_{\alpha} Z_\alpha(k)^{d_\alpha}
\label{eq:Zdynfactorized}
\end{align}
is achieved. The zeta function $Z$ factorizes into zeta functions
belonging to certain irreducible representations $\alpha$ of the
symmetry group. The dimensions $d_\alpha$ of the representations
enter into the full zeta function -- and with them, the quantum
multiplicities of resonances belonging to a certain subspace.

\subsubsection{Assigning weight factors}
The method of cycle expansion expands the zeta function $Z$ into a 
truncated series in which all cycles up to a certain cutoff length 
enter \cite{Cvitanovic1988a,Cvitanovic1989a,Artuso1989a,Wirzba1999a}.
However, besides the primitive cycles, also multiple traversals contribute.
Therefore, it needs to be clarified how repeated revolutions can be taken
into account. Let us assume the primitive fundamental cycles $\tilde{p}$
are known. Then, the contribution of an $r$-times repeated revolution
to the symmetry reduced zeta function \eqref{eq:ZetaSymm} is given by
polynomials such as
\begin{align}
(1 - z^r {t_{\tilde{p},r}})\, ,
\end{align}
where a dummy variable $z$ has been introduced. The cycle weights
$t_{\tilde{p},r}$ have the form of the terms in \eqref{eq:ZGVBilliard} 
and are thus easily calculable from the cycle
weight $t_{\tilde{p}}$ of the primitive fundamental cycle. By using
the factorizations given in Table~\ref {tab:Factorization}, it is
possible to determine the weight factor $w_{\tilde{p}, r}(k)$ as the
sum of all roots ${z_i}^r$ of the polynomials given in the table,
\begin{align}
w_{\tilde{p},r} = \sum\limits_{i} {z_i}^r\, .
\end{align}
If this is possible, a way to use the $\tilde{p}$ for repetitions 
as well has been found. As an example, for the contribution of the 
$r$-times repeated cycle $\tilde{p}$ to the $A_1$ spectrum, we need to solve
\begin{align}
 \left(1-z^r {t_{\tilde{p}, r}}\right)=0 \, ,
\end{align}
which is true for $z^r=1$. Thus, in the $A_1$ subspace, all weight
factors are $w_{\tilde{p},r} = 1$. By this choice, the symmetry
factorization is retained. As another example, consider the $E$
subspace for cycles with $C_3$ symmetry. Here, solutions to the
equation
\begin{align}
 \left(1+ {z^r t_{\tilde{p}, r}} + {z^{r} t_{\tilde{p},r }}^2\right) = 0
\end{align}
are needed. A factorization is given by
\begin{align}
\left(1- \ee^{2\pi\ii r/3} \,t_{\tilde{p},r} 
\right) \left(1- \ee^{-2\pi\ii r/3} \, t_{\tilde{p}, r} \right) = 0\, , 
\end{align}
where the exponentials are the roots $z_i$. Evaluating the sum
${z_1}^r + {z_2}^r$, we find the weight factors $w_{\tilde{p},r} =
-1,-1,2,-1,-1\dots$ for $r=1,2,\dots$. A short notation for this
sequence is given by $w_{\tilde{p},r}=2\cos(2\pi r/3)$. By similar 
calculations, the weight factors $w_{\tilde{p},r}$ given in 
Table~\ref{tab:WeightFactors} are determined.
\begin{table}
\caption{Weight factors $w_{\tilde{p}, r}$ for $r$ traversals of the
primitive cycle $\tilde{p}$. These factors allow for symmetry
factorizations with repetitions of primitive fundamental cycles.}
\begin{tabular}{l| c c c c c}
  & $E$ & $C_3$ & $C_2$ & $S_4$ & $\sigma_d$\\
\hline
  $A_1$ & $1$ & $1$ & $1$ & $1$ & $1$\\[0.6ex]
  $A_2$ & $1$ & $1$ & $1$ & $(-1)^r$ & $(-1)^r$\\[0.6ex]
$E$ & $2$ & $2\cos\frac{2\pi r}{3}$ & $2$ & $1+(-1)^r$ & $1+(-1)^r$\\[0.6ex]
$T_1$ & $3$ & $1+2\cos\frac{2\pi r}{3}$ & $1+2(-1)^r$ &
$1+2\cos\frac{\pi r}{2}$ & $1+2(-1)^r$\\[0.6ex]
$T_2$ & $3$ & $1+2\cos\frac{2\pi r}{3}$ & $1+2(-1)^r$ &
$(-1)^r+2\cos\frac{\pi r}{2}$ & $2+(-1)^r$
\end{tabular}
\label{tab:WeightFactors}
\end{table}

\subsubsection{Ambiguous symmetry}
The shortest cycle labeled by $0$ in the four-sphere system has ambiguous
symmetry. It is possible to map this cycle onto itself by both the
rotation $C_2$ and the reflection $\sigma_d$. This ambiguity requires
special care in the symmetry decomposition. This is particularly
important as the $0$-cycle is one of the fundamental cycles that act
as building block for longer cycles in the sense of cycle expansion. 
The group theoretical weight of the $0$-cycle can be written as
\cite{Cvitanovic1993a}
\begin{align}
{h}_0 = \frac{C_2 + \sigma_d}{2}\, .
\end{align}
The symmetry factorization can thus be not one of those given in
Table~\ref{tab:Factorization}. However, it is possible to use a 
factorization that contains factors in such a way that the factorization 
is at most the greatest common divisor of the factors given for $C_2$ 
and $\sigma_d$ in Table~\ref{tab:Factorization}.
The factorizations and weight factors $w_{0,r}$ 
are given in Table~\ref{tab:WeightsZero}.
\begin{table}
\caption{Factorizations of $Z(k)$ and weight factors $w_{0,r}$ for 
the fundamental cycle $0$ with ambiguous symmetry classes $C_{2}, 
\sigma_{d}$ in all subspaces $\alpha$ of the symmetry group $T_d$.}
\begin{tabular}{l|cc}
   & $C_2$, $\sigma_d$ & $w_{0,r}$ \\
\hline
 $A_1$ & $(1-t_0)$ & $1$ \\
 $A_2$ & $1$ & $0$ \\
 $E$ & $(1-t_0)$ & $1$ \\
 $T_1$ & $(1+t_0)$ & $(-1)^r$ \\
 $T_2$ & $(1+t_0)(1-t_0)$ & $1+(-1)^r$\\
\end{tabular}
\label{tab:WeightsZero}
\end{table}
With that factorization the product \eqref{eq:Zdynfactorized}
of all zeta functions belonging to the irreducible representations of the
symmetry groups coincides with the cycle expansion \eqref{eq:Zdyn}
using all orbits in the full domain.

The final form of the Gutzwiller-Voros zeta function 
we use for our calculations is
\begin{align}
Z_\mathrm{GV;\, \alpha}(k) = \exp\left( - \sum\limits_{\tilde p}\sum\limits_{r=1}^{\infty} 
\frac{1}{r}\frac{w_{\tilde{p},r;\alpha}(-z)^{r n_{\tilde p}}\eto{\ii rl_{\tilde 
p}k}} {\sqrt{\abs{\det( {{\vec{M}}_{\tilde{p}}}^{r} - \identity 
)}}}\right)\, ,
\label{eq:ZGVBilliardSymm}
\end{align}
with $\tilde p$ the primitive symmetry reduced cycles, $r$ the number of
repetitions of it and $\alpha$ the symmetry subspace. A symmetry reduced 
version of the cycle expansion is obtained by expanding 
Eq.\ \eqref{eq:ZGVBilliardSymm} into a power series in $z$ which is 
truncated at a maximum cycle length $n_\mathrm{max}$. Then, $z$ has to be 
set to $z=1$.

\subsection{Harmonic inversion method}
The Gutzwiller-Voros zeta function $Z(k)$ in Eq.\ \eqref{eq:ZGVBilliardSymm}
contains all energy eigenvalues $k$ as complex zeros, and, in principle,
it is possible to obtain spectra by a numerical root search.
This method has been successfully used for billiards, see 
e.g.\ \cite{Wirzba1999a}. However, the root search 
in cycle expansions of high order is numerically expensive. For
statistical purposes it is important not to miss any resonances in 
the strip of the complex plane under consideration. Therefore, a 
dense grid of initial root guesses has to be used for the root 
search. Consequently, many resonances will be found several times. 
Thus, the problem is to distinguish for every new root whether a 
new distinct resonance has been found or if the new zero has already 
been computed. As the number of resonances enters into the fractal 
Weyl law \eqref{eq:WeylLaw} through the counting functions $N(k)$, 
it is crucial to count individual resonances only once.

An alternative to the computation of zeros is based on 
the \emph{harmonic inversion method} for high-resolution spectral 
analysis \cite{Wall1995a,Main99a,Belkic00}.
When Eq.\ \eqref{eq:gk} 
is evaluated along a line of real-valued $k$ or a shifted line 
$k + \ii\delta$ with real $k$ and $\delta$, we obtain a spectrum
\begin{align}
 g(k) = \sum\limits_n \frac{1}{\pi} \frac{\Gamma_n/2
 +\delta}{\left(k-\bar{k}_n\right)^2 + \left(\Gamma_n/2+\delta\right)^2}
\label{eq:LorentzSuperpos2}
\end{align}
which is a superposition of resonances with a Lorentzian shape.
For negative shifts $\delta$ the Lorentzians are located at the positions 
$\bar k_n$, but with reduced widths $\Gamma_n+2\delta$.
The basic idea is now to reformulate, via a Fourier transform, the problem 
of extracting eigenvalues as a signal processing task. 
Details of the method are given in \cite {Wiersig2008a}.

The procedure of calculating quantum spectra is summarized as follows:
First, the spectrum $g(k)$ is calculated as a superposition of 
Lorentzians. We use the cycle-expanded zeta function $Z(k)$ for this 
purpose. The quantity 
\begin{align}
g(k) = -\frac{1}{\pi}\Imag\firstderiv{}{k} \ln Z(k) = 
-\frac{1}{\pi}\Imag\frac{Z'(k)}{Z(k)}
\label{eq:gTildeHI}
\end{align} is evaluated along lines parallel to the real axis with 
different shifts $\delta$. Thus, the shifts that allow for better 
results in harmonic inversion enter into the cycle expansion. Then, 
harmonic inversion is used to obtain the Lorentzians' parameters 
$\bar k_n$ and $\Gamma_n$ for spectra calculated with different 
shifts. In the next step, the spectra are filtered via the 
amplitudes. The quantity $\tilde{g}(k)$ given in \eqref{eq:gTildeHI} 
should give resonances with an amplitude of $A_n = 1$. 
True resonances with amplitudes $A_n\approx 1$ can be clearly separated
from spurious resonances with nearly zero amplitudes.
Finally, the spectra for different shifts $\delta$ are joined such 
that the individual strips do not overlap.

\section{Classical escape rate}
\label{sec:escape}
The classical escape rate $\gamma_0$ can be interpreted descriptively as 
follows \cite{Gaspard1989a}: presume the scattering system under consideration 
is located in a box much larger than the system itself. Conducting $N_0$ 
scattering experiments with the same incident energy $k$, but different 
incident directions, one finds that the number $N_t$ of trajectories that 
are inside the box after the time $t$ has passed decays exponentially as
\begin{align}
N_t \propto N_0 \eto{-\gamma_0 t}\, .
\label{eq:classDec}
\end{align}
The relation of the escape rate and the imaginary part of the quantum 
resonances can be understood from the correspondence principle. The number of 
classical trajectories inside the box corresponds to the quantum 
probability density $\qmnorm{\psi}$. The decay of this probability,
\begin{align}
\qmnorm{\psi} \propto \eto{-\Gamma t}\, ,
\end{align}
relates to the decay \eqref{eq:classDec} of the number of classical 
trajectories inside the box. Thus, in the classical limit
\begin{align}
\Imag k_n = -\frac{\Gamma_n}{2} \to -\frac{\gamma_0}{2}
\end{align}
holds.

The classical escape rate can be calculated by the method of cycle expansion 
as well \cite{Cvitanovic1989a}. The escape rate $\gamma_0$ is found to 
be the largest real zero of a dynamical zeta function
\begin{align}
Z(s) = \prod\limits_{\tilde p} (1-t_{\tilde p}(s)) \, ,
\end{align}
with $\tilde p$ the primitive periodic cycles and $t_{\tilde p}$ the cycle weights. 
For a three-dimensional system, 
\begin{align}
t_{\tilde p}(s) = \frac{\eto{-l_{\tilde p} s}}{\abs{\lambda_{\tilde p}^{(1)}\lambda_{\tilde 
p}^{(2)}}}  \, .
\end{align}
The quantities $\lambda_{\tilde p}^{(i)}$ are the leading stability 
eigenvalues. A generalization to a zeta function for three 
dimensions and multiple traversals $r$ of the primitive cycle $\tilde
p$ is given by
\begin{align}
Z(s) = \exp\left( -\sum\limits_{\tilde p} \sum\limits_{r} 
\frac{1}{r}\frac{\eto{-r l_{\tilde p} s}}{\abs{\lambda_{\tilde p}^{(1)} 
\lambda_{\tilde p}^{(2)}}^r} \right) \, ,
\label{eq:GutzEsc}
\end{align}
with $\lambda_{\tilde p}^{(i)}$ the leading two stability 
eigenvalues of $\tilde p$. This zeta function can be cycle-expanded 
as described in Sec.\ \ref{chap:Semiclass}. Results for the escape 
rate $\gamma_0$ at various configurations $d/R$ are given in Table~
\ref{tab:escrate}.

\begin{table}
\caption{Classical escape rates $\gamma_0^{(n)}$ in order $n$ of the cycle 
expansion for various values of the configuration parameter $d/R$.}
\begin{tabular}{c| c c c c}
$d/R$ & $\gamma_0^{(1)}$ & $\gamma_0^{(2)}$ & $\gamma_0^{(3)}$ & $\gamma_0^{(4)}$\\
\hline
$4$ & $1.16655$ & $1.16459$ & $1.16440$ & $1.16440$\\
$6$ & $0.85042$ & $0.84977$ & $0.84974$ & $0.84974$ \\
$8$ & $0.68259$ & $0.68230$ & $0.68230$ & $0.68230$ \\
$10$ & $0.57634$ & $0.57619$ & $0.57619$ & $0.57619$
\end{tabular}
\label{tab:escrate}
\end{table}

\section{Results}
\label{chap:Results}
The fractal Weyl law has been put to test for billiard systems 
before. In \cite{Lu2003a}, the 3-disk billiard has been studied. To 
make our own results comparable to those given in \cite{Lu2003a}, we 
carry out a similiar discussion.

\subsection{Defining a scale for the strip widths}
For the 3-disk system discussed in \cite{Lu2003a}, the strip widths 
$C$ have been chosen in relation to the classical escape rate 
$\gamma_0$. For large values $k\to\infty$, the imaginary part of 
quantum resonances converges to $\Imag k = -\gamma_0/2$ 
\cite{Lu2003a, Pance2000a}. Thus, the discussion of the results is 
simplified by rescaling the strip widths $C$ to 
\begin{align}
\tilde C := \frac{C}{\gamma_0/2}\, ,
\end{align}
which defines a universal scale independent of the symmetry subspace 
and the ratio $d/R$. Similar to \cite{Lu2003a}, we evaluate the 
fractal Weyl law for scaled strip widths $\tilde C \in [1; 1.6]$.

\subsection{Counting resonances}
We have computed spectra for various values of $d/R$ in all symmetry 
subspaces. Generally, we find the best convergence behavior of cycle 
expansions for large values of $d/R$. Furthermore, the 
one-dimensional group representations $A_{1}$ and $A_{2}$ yield the 
largest number of converged resonances. The two-dimensional 
representation $E$ and the three-dimensional $T_{1}$, $T_{2}$ 
representations converge not as well in cycle expansion since the 
shadowing of individual cycles is less efficient for the weight 
factors of these subspaces. For $A_{1}$, where all weight factors 
are equal to $1$, the best convergence is observed.

It is important to note that for the tests of the fractal Weyl law we have 
used only {\em converged} resonances.
For example, the $A_1$ resonances at $d/R=10$ are sufficiently converged in 
the region $\Real k < 6000$, $\Imag k > -0.45$ so that counting functions 
$N(k)$ obtained in orders 10 and 11 of the cycle expansion fully agree.
Thus, comparisons with the fractal Weyl law are not affected by the order 
of the cycle expansion.

\begin{figure}
\includegraphics[width=0.95\columnwidth]{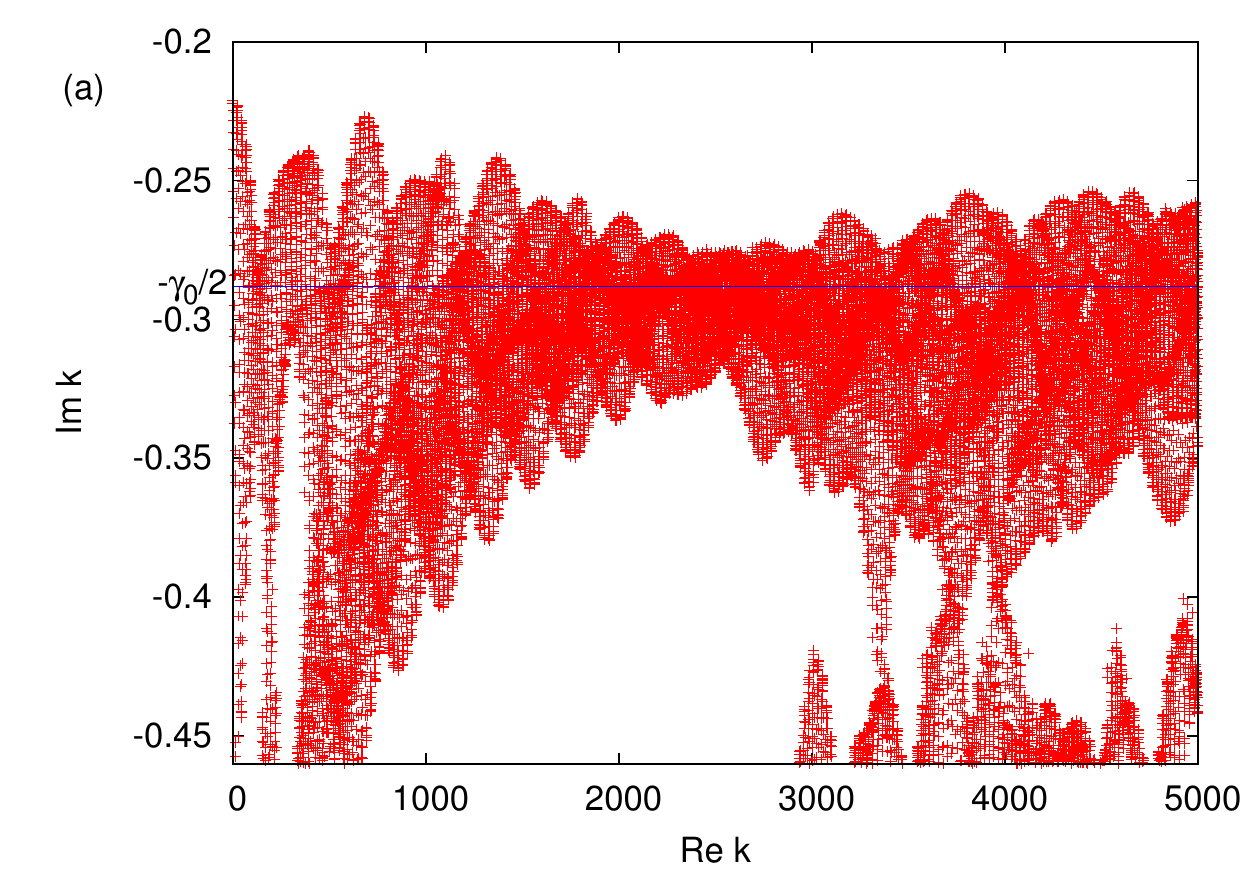}
\includegraphics[width=0.95\columnwidth]{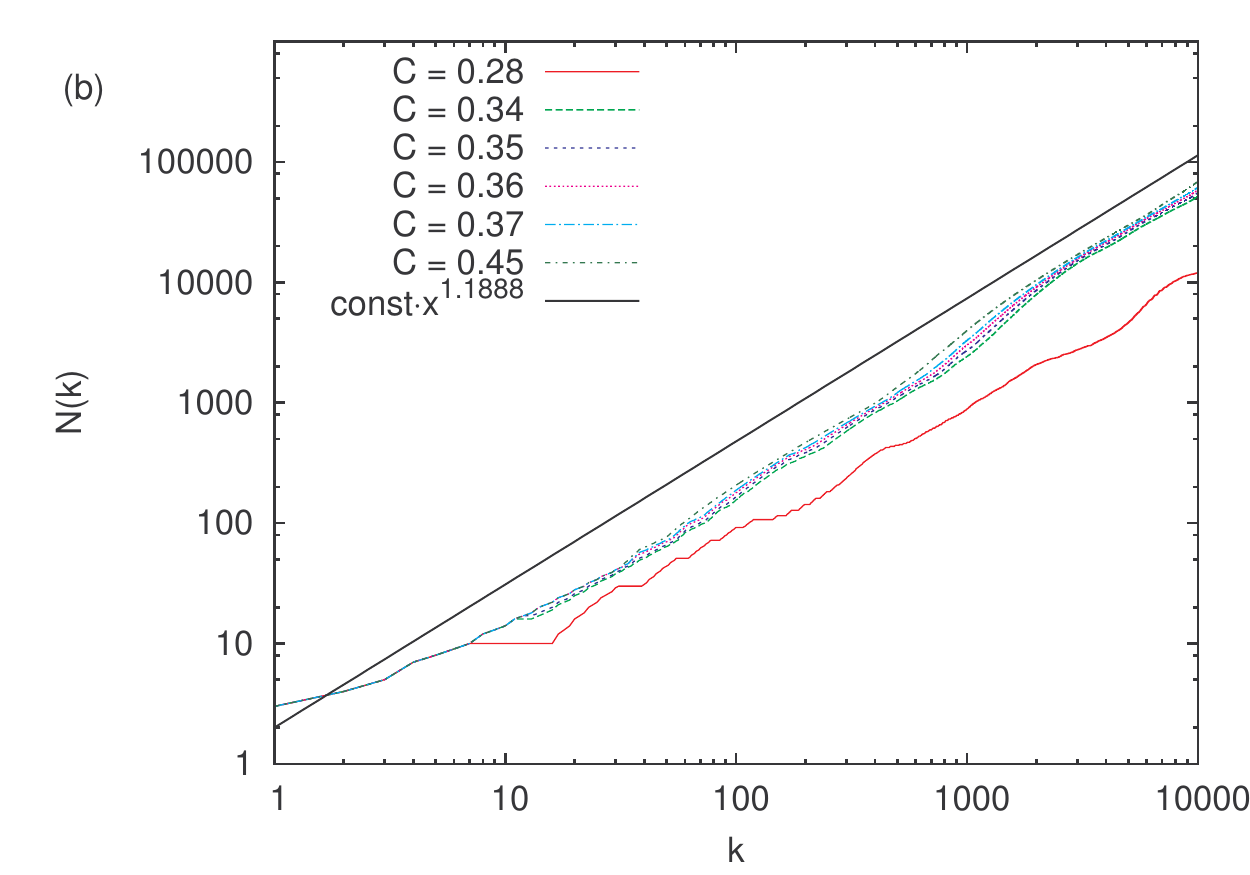}
\caption{
(a) Spectrum and (b) counting functions $N(k)$ for $A_1$ 
resonances calculated from cycle expansion in order $11$ for the 
ratio $d/R=10$. Several counting functions for different strips $C$ 
are shown. The curves can be used as ``raw data'' to fit power laws.
In this way, the exponent $\alpha$ in the fractal Weyl law can be obtained 
and compared to the classical calculations. More 
than $50\,000$ resonances have been used in the analysis.}
\label{fig:CountFuncd10}
\end{figure}

The fractal Weyl law \eqref{eq:WeylLaw} suggests that the counting 
functions $N(k)$ obey a power law,
\begin{align}
N(k) \propto k^\alpha\, ,
\label{eq:PowerLaw666}
\end{align}
thus, in a logarithmic plot of $N(k)$, straight lines of slope 
$\alpha=1+\dH$ 
are expected. A sample spectrum and corresponding counting functions 
for $d/R=10$ are shown in Fig.~\ref{fig:CountFuncd10}.
This figure is generic in structure, i.e.\ we have found similar 
behavior of $N(k)$ in other subspaces and for other ratios $d/R$ as 
well. Thus, a brief discussion of these features will be given in 
the following.

We first note that the strip width $C$ has to be sufficiently large, 
since otherwise, counting would not involve reasonably large 
numbers of resonances. In the spectrum shown in Fig.\ \ref
{fig:CountFuncd10}(a), we have found converged resonances in the relevant 
strip $\tilde C \in [1; 1.6]$ for $\Real k \lesssim 6000$. For small 
strip widths such as $C=0.28 \Leftrightarrow \tilde C = 0.97$, it is 
evident from Fig.~\ref{fig:CountFuncd10}(b) that counting involves only 
a limited number of resonances. Larger strip widths involve more 
resonances in the counting. However, choosing the strip width too 
large, the counting may also involve resonances that may not have 
been converged. Taking the asymptotic behavior of the resonances' 
imaginary parts into account, choosing rescaled strip widths in the 
interval $\tilde C \in [1.0; 1.6]$ turns out to be a reasonable choice.

Figure~\ref{fig:CountFuncd10}(b) reveals that the counting functions 
$N(k)$ deviate from power laws that led to straight lines in the 
plot. From this observation one infers that the exponent $\alpha$ 
will clearly depend on the $k$-range one fits to. We follow \cite
{Lu2003a} and choose the largest interval converged resonances have 
been computed for.

\subsection{Putting the fractal Weyl law to test}
\label{sec:PutTest}
To provide several tests for the fractal Weyl law, we 
will provide plots showing the exponents $\alpha$ obtained from 
least-squares fitting a power function $N(k) \propto k^\alpha $ to 
the counting functions calculated from the spectra for various 
subspaces, fitting ranges $\left[0;\, k_\mathrm{max}\right]$, 
configuration parameters $d/R$ and strip widths $\tilde C$. We have 
performed least-squares fits to match the power function \eqref
{eq:PowerLaw666} to the measured $N(k)$.

\subsubsection{Configuration $d/R=6$}
\begin{figure}
\includegraphics[width=0.9\columnwidth]{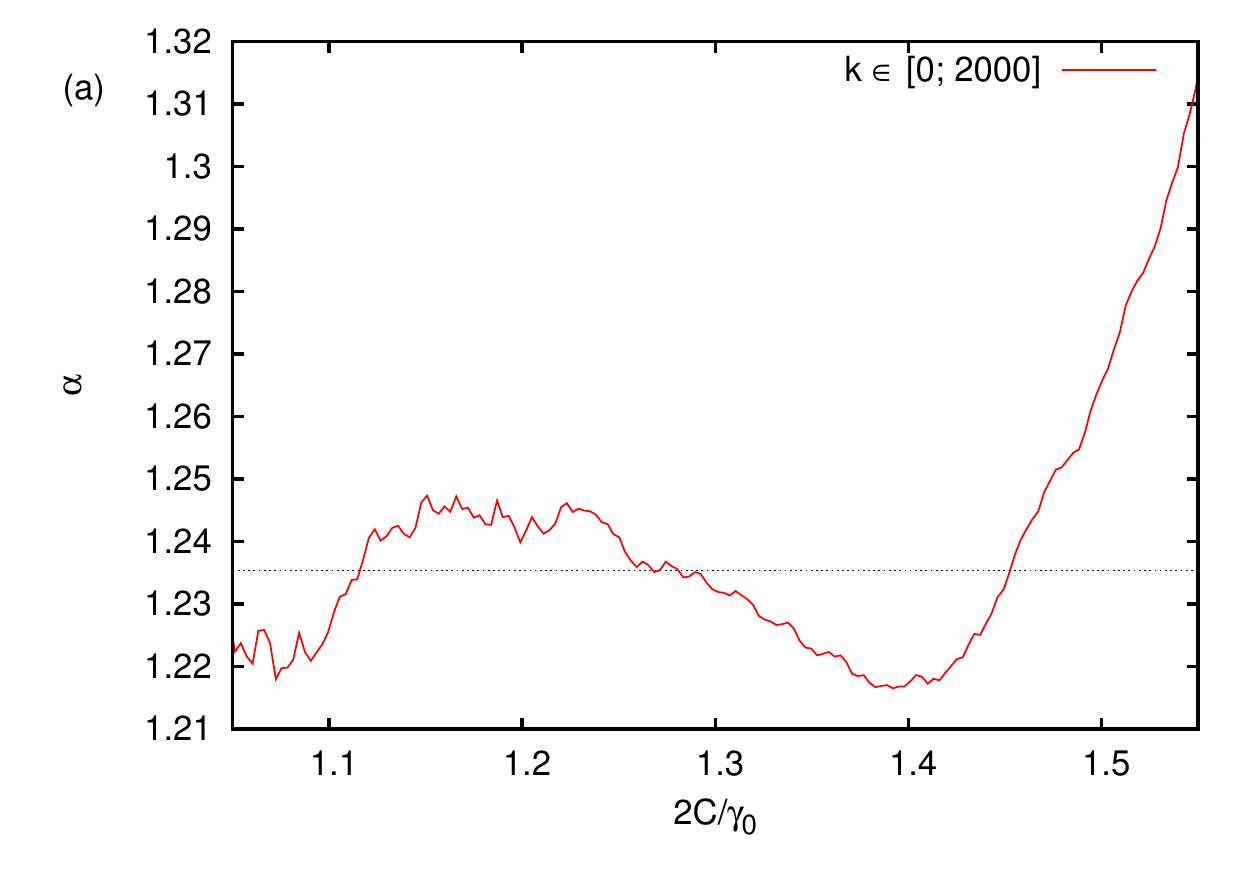}
\includegraphics[width=0.9\columnwidth]{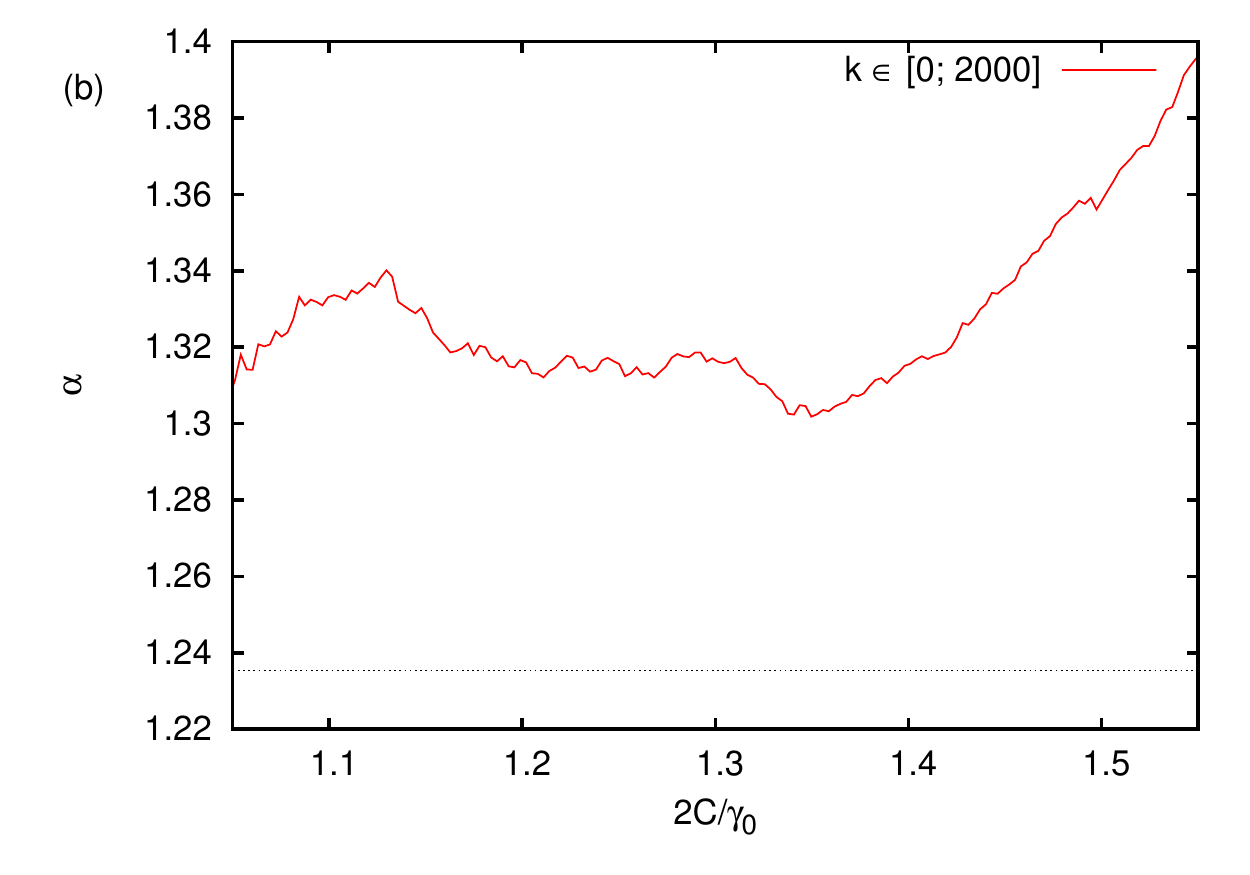}
\caption{
Exponents $\alpha$ obtained from least-squares fits of a 
power law to measured counting functions for (a) $A_1$ resonances and 
(b) $A_2$ resonances calculated for $d/R=6$ in order 13 of the cycle expansion. 
The power law has been fitted to the interval $k \in [0; 2000]$. The 
vertical dotted line gives the classical exponent $\alpha=1 + \dH = 1.2354$.}
\label{fig:expd6}
\end{figure}
For $d/R=6$, we obtain the exponents shown in Fig.~\ref{fig:expd6}. For the 
$A_1$ subspace, we find a very good agreement for moderate values of 
$\tilde C < 1.4$. The relative error in this $\tilde C$-interval is 
less than 2 percent. However, in the $A_2$ subspace, all computed 
exponents are too large by about 8 percent for the same $\tilde 
C$-interval. One possible reason is that the $k$-range used for 
fitting is too small.

\subsubsection{Configuration $d/R=8$}
\begin{figure}
\includegraphics[width=0.9\columnwidth]{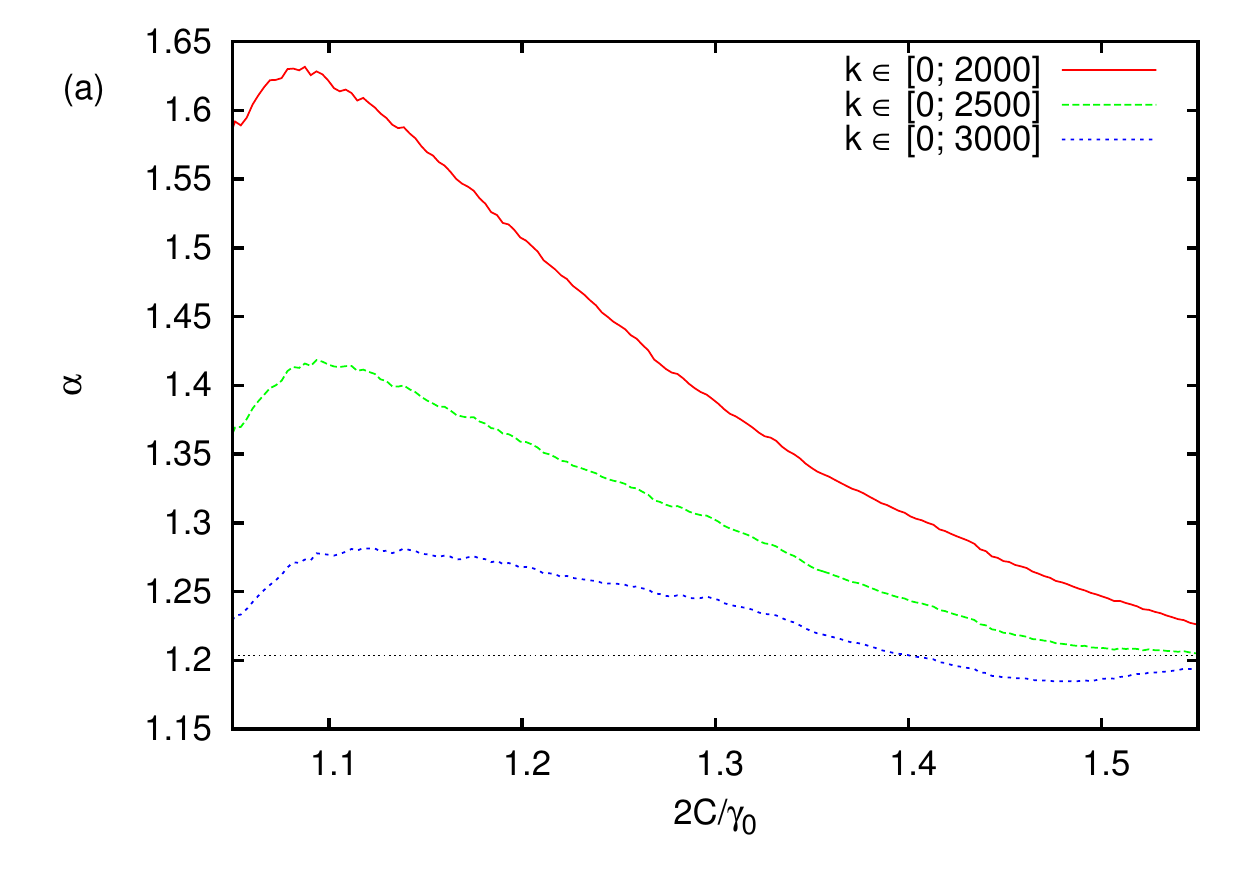}
\includegraphics[width=0.9\columnwidth]{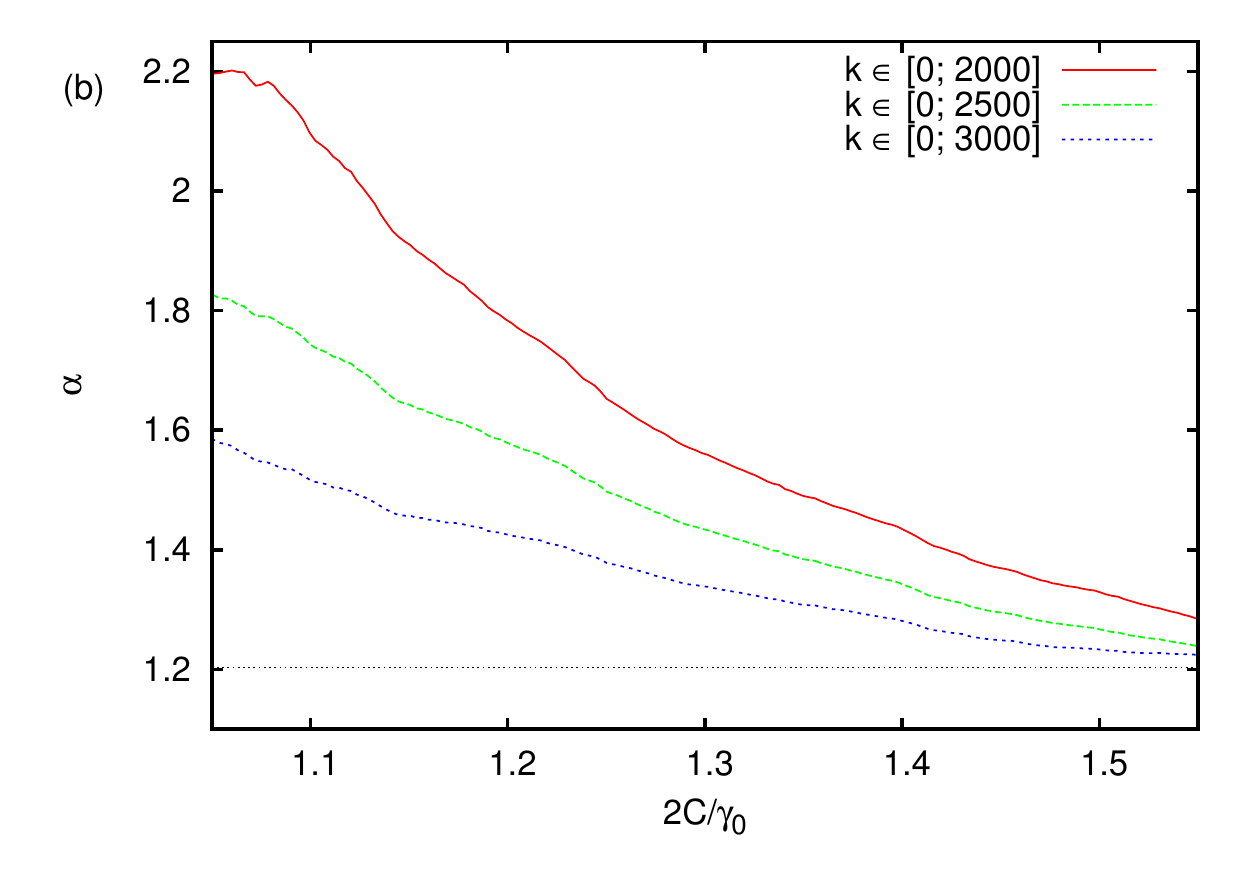}
\caption{
Exponents $\alpha$ obtained for $d/R=8$ from least-squares fits of a 
power law to measured counting functions for (a) $A_1$ resonances and (b)
$A_2$ resonances. The power law has been fitted to several $k$-intervals. 
The vertical dotted line gives the classical exponent $\alpha=1 + \dH = 1.2063$.}
\label{fig:expd8}
\end{figure}

Performing the same procedure for a configuration parameter of 
$d/R=8$, we obtain the plots shown in Fig.~\ref{fig:expd8}.

Both plots reveal a clear tendency to obey the fractal Weyl law 
within a smaller error range when the range of $k$ values used for the 
fit increases. However, for reasons of convergence, longer spectra 
have not been used. We note that for $\tilde C < 1.4$ and $k \in [0; 
3000]$, the error is less than 7 percent for the $A_1$ 
resonances. The exponents obtained from $A_2$ resonances are larger 
than the expected exponent. For $\tilde C=1.3$, the relative error 
is about 15 percent.

\subsubsection{Configuration $d/R=10$}
\begin{figure}
\includegraphics[width=0.9\columnwidth]{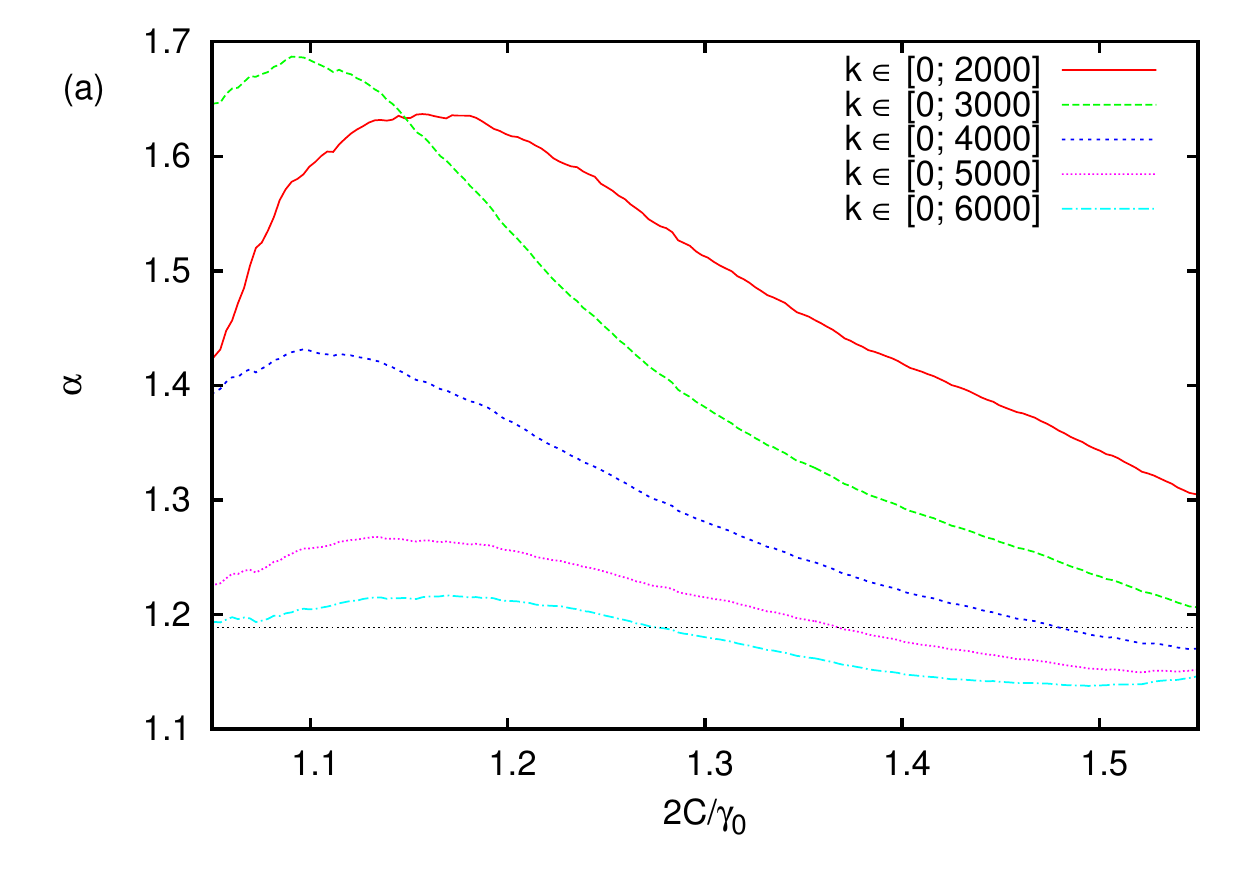}
\includegraphics[width=0.9\columnwidth]{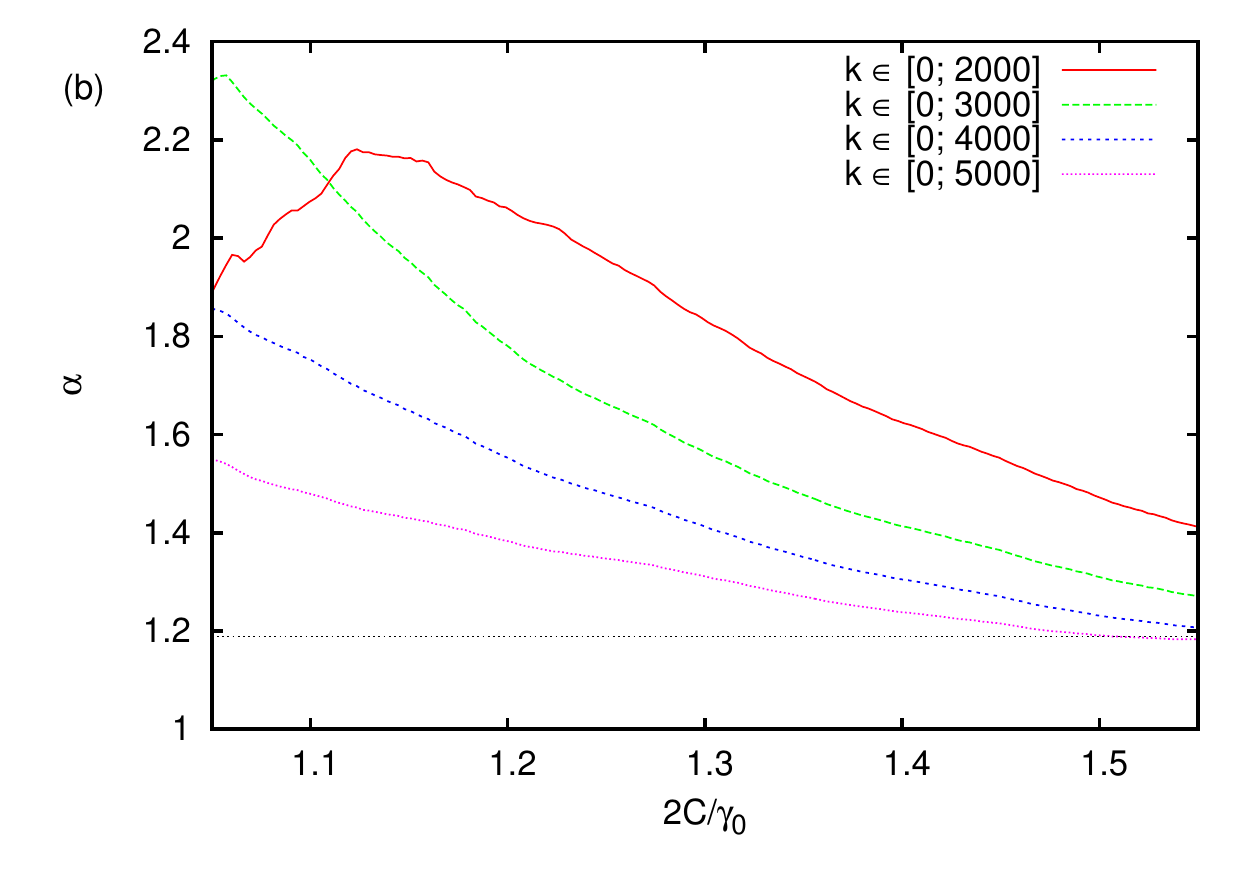}
\caption{
Exponents $\alpha$ obtained for $d/R=10$ from least-squares fits of a 
power law to measured counting functions for (a) $A_1$ resonances and (b)
$A_2$ resonances. Results for several $k$-intervals are shown. The 
vertical dotted line gives the classical exponent $\alpha=1 + \dH = 1.1888$.}
\label{fig:expd10}
\end{figure}

Finally, the system configuration given by $d/R=10$ has been studied. 
Results are shown in Fig.~\ref{fig:expd10}.

The counting functions for the $A_1$ subspace once more tend to give 
the expected exponent when the $k$-interval used for the fit 
increases. For $k\in [0; 6000]$, the error is less than 3 percent. 
Again, the $A_2$ spectra yield exponents that are too large. 
Possibly the $k$-range investigated here is not large enough to 
exhibit the asymptotic behavior clearly.

\subsection{Discussion}
While the classical calculations for the fractal dimension $\dH$ are 
accurate to at least four significant digits, the agreement of the 
exponents $\alpha$ is at best $2$ to 7 percent for the $A_{1}$ 
spectra. However, we note that for the two-dimensional three-disk 
billiard, the errors in the exponents are about $5$ to 
10 percent \cite{Lu2003a}. Therefore, we conclude that the 
fractal Weyl law for the four-sphere scattering system is confirmed 
with roughly the same accuracy as for the three-disk billiard. 

A very large $k$-range seems to be necessary for a proper 
investigation. This tendency is also visible for the $A_2$ spectra. 
The exponents obtained from the $A_2$ spectra are too large. 
However, using larger $k$-intervals, the exponents seem to approach 
the correct value for large strip widths $\tilde C$. Possibly, if 
larger spectra were available, the expected exponents could be 
obtained. Unfortunately, we are limited by the convergence of the 
cycle expansions we use. The higher-dimensional symmetry subspaces 
$E$, $T_1$ and $T_2$ could not be used to put the fractal Weyl to 
test law since the spectra did not contain enough converged 
resonances.

\section{Summary and outlook}
\label{sec:conclusion}
This paper provides a test of the fractal Weyl law for a 
three-dimensional scattering system. The four-sphere billiard was 
investigated both classically and quantum mechanically.

In Sec.~\ref{chap:Gauging}, we have developed a fast and very 
precise method to gauge the repeller. We found estimates for the 
Hausdorff dimension $\dH$ with a relative accuracy of $10^{-4}$. 
Although the algorithm is based on strong assumptions, it works over 
a wider range of the configuration parameter $d/R$ than existing 
methods.

In Sec.~\ref{chap:Semiclass}, we have discussed the methods of
semiclassical quantization. We have applied the method of cycle
expansion to the four-sphere billiard. Furthermore, for the first
time, the method of symmetry decomposition was demonstrated for the
Gutzwiller-Voros zeta function of the system.

We have given results in Sec.~\ref{chap:Results}. We have provided 
tests of the fractal Weyl law for various configurations of the 
system. Although we have found the counting functions $N(k)$ to 
deviate from power functions, we could confirm the fractal Weyl law 
for the $A_{1}$ resonances of the four-sphere scattering within a 
small error range. For those spectra we did not find a 
convincing agreement of calculated level numbers $N(k)$ with the 
prediction $N(k) \propto k^{1+\dH}$ for, there is hope that larger 
spectra would lead to the expected exponent. We also assume that the 
deviations from pure power laws are due to the fact that the energy 
range under consideration is too small.

As an outlook, the physical origin of the modulations in the
counting functions $N(k)$ will have to be investigated. Moreover, it
is desirable to study further three-dimensional scattering systems
to find out to what extent the results found for the four-sphere billiard
carry over and are generic.

%

\end{document}